\documentclass[runningheads]{llncs}

\usepackage[english]{babel}

\usepackage{cite}

\usepackage[english]{babel}

\usepackage{amsmath}
\usepackage{amssymb}
\usepackage{graphicx}

\usepackage[colorlinks=true, allcolors=blue]{hyperref}

\usepackage{algorithm}
\usepackage{algorithmicx}
\usepackage{algpseudocode}

\usepackage{xcolor}
\usepackage[final]{changes}
 
\definechangesauthor[name=Feedback, color=olive]{F}
\definechangesauthor[name=Review01, color=blue]{R1}
\definechangesauthor[name=Review02, color=green]{R02}
\definechangesauthor[name=Review03, color=red]{R03}
\definechangesauthor[name=Review04, color=purple]{R04}

\title{Compiler support for semi-manual AoS-to-SoA conversions with data views}
\titlerunning{Semi-manual AoS-to-SoA conversions with data views}

\author{Pawel K.~Radtke\inst{1}\orcidID{0009-0009-8613-3632} \and Tobias
Weinzierl\inst{1}\orcidID{0000-0002-6208-1841}}

\authorrunning{P. Radtke et T. Weinzierl}

\institute{Department of Computer Science, Durham University, Durham, United
Kingdom \email{\{pawel.k.radtke,tobias.weinzierl\}@durham.ac.uk}\\
\url{https://scicomp.webspace.durham.ac.uk}}

\begin{document}
\maketitle

\begin{abstract}
  The C programming language and its cousins such as C++ stipulate the static
storage of sets of structured data:
Developers have to commit to one, invariant data model---typically a
structure-of-arrays (SoA) or an array-of-structs (AoS)---unless they manually
rearrange, i.e.~convert it throughout the computation.
Whether AoS or SoA is favourable depends on the execution context and
algorithm step.
We propose a language extension based upon C++ attributes through which
developers can guide the compiler what memory arrangements are to be used.
The compiler can then automatically convert (parts of) the data into the format
of choice prior to a calculation and convert results back afterwards.
As all conversions are merely annotations, it is straightforward for the
developer to experiment with different storage formats and to pick subsets of
data that are subject to memory rearrangements.
Our work implements the annotations within Clang and
demonstrates their potential impact through a smoothed particle hydrodynamics (SPH) code.

\end{abstract}

\begin{keywords}
  Array-of-structs, struct-of-arrays, memory layout transformations, data views,
  compiler, vectorisation
\end{keywords}

\section{Introduction}
\label{section:introduction}

%
%
Loops over sequences of data are the workhorses of scientific
codes.
Modern high-level languages provide us with the concept of structures 
to model our data elements.
They are convenient to represent particles, mesh cells, and so
forth.
Due to its \texttt{struct}s, the C++ language leans towards an
array-of-structs (AoS) storage for sequences.
The class is the primary modelling entity for the programmer, and the
language favours sequences over class
instances,
i.e.~objects~\cite{Hirzel:2007:DataLayoutForOO,Homann:2018:SoAx,Reinders:2016:XeonPhi,Jubertie:2018:DataLayoutAbstractionLayers}.

%
%
In many cases, implementations over structure-of-arrays (SoA) 
outperform their SoA counterpart.
They facilitate the efficient usage of vector
instructions~\cite{Gallard:2020:ExaHyPEVectorisation,Intel:MemoryLayoutTransformations,Springer:2018:SoALayout}
and are less sensitive to cache effects
\cite{Homann:2018:SoAx,Hundt:2006:StructureLayoutOptimisation,Springer:2018:SoALayout}:
With AoS, compute kernels over sequences of structs 
have to gather and scatter vector register content (shuffling),
vector parallelism is not blatant to the compiler, 
and structs for multiple loop iterations might not fit into the cache.
With SoA, vector registers can be filled with coalesced memory
accesses, vector computations are exposed explicitly, and data from subsequent
loop iterations is likely to reside in cache.
While CPUs improve their gather-scatter efficiently with every new generation,
the observations remain valid and apply
GPUs, too~\cite{Strzodka:2011:AbstractionSoA}.

SoA is not always superior to AoS. 
Tasks such as boundary data exchange, particle movements over meshes, or, in
general, any algorithm that has to alter or permute struct sequences or
access it in a non-continuous way \cite{Strzodka:2011:AbstractionSoA} benefit
from AoS.
The size of characteristic sequences \cite{Homann:2018:SoAx} and the memory
footprint per struct further affect which storage format performs better.
Finally, any runtime difference depends upon how successful the compiler 
vectorises an algorithm
\cite{Jubertie:2018:DataLayoutAbstractionLayers,Xu:2014:SemiAutomaticComposition} and what the target architecture looks like.
The choice of an optimal data structure is context-dependent.
There is no ``one format rules them all''.

%
%
Refactoring code to accommodate memory rearrangements is
error-prone and laborious.
Wrappers allow developers to write their
algorithms in a memory layout-agnostic way.
C++ template meta programming 
combined with specialised containers are
popular to achieve this
\cite{Homann:2018:SoAx,Reinders:2016:XeonPhi,Jubertie:2018:DataLayoutAbstractionLayers,Springer:2018:SoALayout,Strzodka:2011:AbstractionSoA}.
This is a \emph{static, global} approach. 
The data layout remains fixed.
\added[id=R02]{
 Any static strategy fails to react to the algorithmic context.
 A \emph{dynamic, local} approach reorders data prior to the loop, such that only a local
 code block is aggressively optimised while the global data layout stays
 invariant.
}

To determine the permutation of data relative to plain AoS, 
we distinguish \emph{manual} (user-driven) from \emph{automatic} workflows.
A \emph{guided} one
\cite{Jubertie:2018:DataLayoutAbstractionLayers,Xu:2014:SemiAutomaticComposition}
is a hybrid, where the actual transformation is 
\replaced[id=R04]{carried out by the compiler, but only at the explicit request
of the user}{done by a compiler middleman, yet manually steered by the user}.
%
%
\replaced[id=R02]{We}{Any static strategy fails to react to the algorithmic
context. Therefore, we}  propose a \emph{dynamic}\replaced[id=R02]{, local,
guided approach}{ alternative}:
The code remains written in plain C++.
Through additional annotations, programmers specify alternative data layouts
for particular loops.
A compiler takes the annotations and manually reorders data in
a separate, temporary memory block prior to the loop.
This is an out-of-place reordering, i.e.~does not alter the original
data layout
\cite{Gallard:2020:ExaHyPEVectorisation,Gustavson:2012:InPlaceStorage,Vikram:2014:LLVM}.
The compiler also alters all corresponding data accesses such that they fit
to the reordered copy of the data.
A counterpart annotation ensures that data modifications are copied back at the end of the
block, i.e.~data are kept consistent.

%
%
Our approach is lightweight and non-invasive:
The code remains correct if the annotations are unknown to the compiler.
\replaced[id=R02]{
 Different to manual optimisations or the implementation through future C++
 generations supporting reflections~\cite{Childers:2024:CPPReflections},
 no
}{No}
code rewrites are required.
This facilitates experimenting with different data layouts and separating
algorithm development from performance tuning
\cite{Gallard:2020:Roles}.
Our approach is not lightweight behind the scenes, as it introduces data
movements.
To reduce data movements, we rely on \emph{views}:
Developers can apply the AoS-to-SoA and SoA-to-AoS permutations to a subset of
the structs' attributes.

We demonstrate the potential of the idea by means of selected SPH
kernels~\cite{Schaller:2024:Swift}.
Individual SPH interactions can either be strongly memory-bound, or
rely on compute-intense kernels.
Since we allow the code base to stick to AoS overall, we do not negatively
impact algorithmic phases such as the particle boundary exchange or any sorting,
but still show that some kernels---depending on the context and the algorithmic
character---perform better due to temporal data reordering.
We cannot yet provide a heuristic when conversions pay off. 
Our data
suggests that common knowledge that it pays off only for Stream-like \cite{McCalpin:1995:Stream} kernels
\cite{Homann:2018:SoAx} or large arrays \cite{Strzodka:2011:AbstractionSoA} is
not unconditionally true.

Dynamic data structure transformations within a code have been used by
several groups
successfully \cite{Gallard:2020:ExaHyPEVectorisation,Vikram:2014:LLVM}.
Our contribution is that we clearly separate storage format considerations from
programming, introduce the notion of views, and move all data conversions
into the compiler.
As this approach requires no rewrite of existing code\replaced[id=F]{,}{ or}
additional libraries
\added[id=R02]{ or new features such as
reflections~\cite{Childers:2024:CPPReflections}}, it has the potential to streamline code development and code tuning.
\added[id=R02]{We notably do not increase the syntactic complexity of the code.}
As we challenge the common knowledge that temporal data reordering
prior to loops or computational kernels\replaced[id=R1]{is}{, in general,}
problematic~\cite{Hundt:2006:StructureLayoutOptimisation,Intel:MemoryLayoutTransformations},
we lay the foundations of making these data layout optimisations a standard
optimisation step within a compiler pipeline.

%
%
The remainder is organised as follows:
We sketch our use case first to motivate our work
(Section~\ref{section:demonstrator}), though all ideas are more widely applicable.
We next introduce our code annotations, and
then discuss their semantics (Section~\ref{section:annotations}).
This allows us to realise the annotations in
Section~\ref{section:realisation}, before we finally study their impact.
A brief discussion and outlook in Section~\ref{section:conclusion} close the
discussion.

\section{Demonstrator use case}
\label{section:demonstrator}

We motivate and illustrate our ideas by means of the compute kernels of a
simple smoothed particle hydrodynamics (SPH) code.
SPH is a Lagrangian technique.
The particles interact, move and carry properties.
This way, they represent the underlying fluid flow.
We zoom into the elementary operations of any SPH code, which is the
hydrodynamic evolution with leapfrog as SPH's predominant time integrator.
More complex physical models typically start from there \cite{Schaller:2024:Swift}.

\paragraph{Algorithmic kernels.}


\replaced[id=R04]{We}{W.l.g., we} focus on $N$ particles and assume that they can interact with any 
other particle.
This resembles an SPH code which clusters the computational domain into control
volumes each hosting a small, finite number of particles.
The control volumes are chosen such that particles only interact with other
particles within the volume and its neighbours, as SPH typically works 
with a finite interaction radius.
We assume to know its upper bound.
The core compute steps of the algorithm then read as follows:

\begin{enumerate}
  \item We determine the smoothing length of each particle, i.e.~the
  interaction area, and its \emph{density}. This initial step evaluates
  two-body interactions, i.e.~is in $\mathcal{O}(N^2)$. It studies all
  nearby particles within the volume and decides if to shrink
  or increase the interaction radius. The process then repeats. The
  density computation overall is iterative, but we study one
  iteration step only.
  \item We compute the \emph{force} acting on each particle. The force
  is the sum over all forces between a
  particle and its neighbours within the interaction radius.
  Overall, the force calculation is asymptotically of quadratic complexity.
  \item We \emph{kick} the particle by half a time step, i.e.~we accelerate it.
  This is a mere loop over all particles within a cell and hence of linear
  complexity.
  \item We \emph{drift} a particle, i.e.~update its position. This is another
  loop in $\mathcal{O}(N)$.
  \item We \emph{kick} a second time, i.e.~add
  another acceleration.
\end{enumerate}

\paragraph{Memory layout and data flow.}

Our SPH particle is modelled as a struct.
We work with AoS.
The hosted structs aka 
particles can hold from a few quantities up to hundreds of doubles.
Our benchmark code induces a memory footprint of 256 bytes per particle.
Some of these bytes encode administration information, others store physical
properties.
For many steps, only few attributes enter the equations.
The others are, within this context, overhead
\cite{Bungartz:2010:Precompiler,Homann:2018:SoAx}.

The density calculation starts from the density and smoothing length of the
previous time step and updates those two quantities, and others such as the neighbour count, the rotational velocity vector, and various derivative properties.
Smoothing length, density and further properties feed into the force
accumulation which eventually yields an acceleration per particle.
Kicks are relatively simple, i.e.~add scaled accelerations onto the velocity.
The second kick in our implementation also resets a set of predicted values
which feed into the subsequent density calculation.
Drifts finally implement a simple Euler time integration step, i.e.~a tiny
daxpy.

SPH is a Lagrangian method and hence works with particles which are scattered
over the computational domain.
While we tend to hold the particles continuously in memory per cell to facilitate
vectorisation, the nature of the moving particles implies that we have to resort
frequently.
In a parallel domain decomposition environment, we furthermore have to exchange
few particles per time step as part of the halo exchange.


\section{Source code annotations}
\label{section:annotations}

Converting AoS into SoA is an evergreen in high-performance computing,
once we have committed to AoS as development data structure (cmp.~for example
\cite{Gallard:2020:Roles,Gallard:2020:ExaHyPEVectorisation,Homann:2018:SoAx,Hundt:2006:StructureLayoutOptimisation,Intel:MemoryLayoutTransformations,Jubertie:2018:DataLayoutAbstractionLayers,Reinders:2016:XeonPhi,Springer:2018:SoALayout,Strzodka:2011:AbstractionSoA,Sung:2012:DataLayoutTransformations,Vikram:2014:LLVM,Xu:2014:SemiAutomaticComposition}).
Vector computing in a SIMD or SIMT sense including coalesced memory accesses,
cache and TLB effects drives such rewrites.
A sophisticated conversion takes into account weather we have to convert only
a subset of the struct, i.e.~if we can peel a struct or open a view
\cite{Hundt:2006:StructureLayoutOptimisation}.


\begin{algorithm}[htb]
  \begin{algorithmic}[1]
 \Function{void drift}{Particle *particles,  int size}
 \State [[clang::soa\_conversion\_target(particles)]]
 \State [[clang::soa\_conversion\_target\_size(size)]]
 \State [[clang::soa\_conversion\_inputs(pos, vel, updated)]]
 \State [[clang::soa\_conversion\_outputs(pos, updated)]]
 \For{(int i = 0; i $<$ size; i++)}
   \State auto \&p = particles[i];
   \State p.pos[0] += p.vel[0] * dt;
   \State p.pos[1] += p.vel[1] * dt;
   \State \ldots
   \State p.updated = true;
 \EndFor
 \EndFunction
\end{algorithmic}

  \caption{
    The drift of all particles within a control volume is annotated with
    instructions to the compiler to convert parts of the underlying AoS data
    into SoA prior to the actual loop invocation.
    \label{algorithm:annotations:example}
    }
\end{algorithm}

We propose that developer focus exclusively on either SoA or AoS.
For the demonstrator from Section~\ref{section:demonstrator}, 
AoS is, in line with many
codes~\cite{Hirzel:2007:DataLayoutForOO,Homann:2018:SoAx,Reinders:2016:XeonPhi,Jubertie:2018:DataLayoutAbstractionLayers}, 
a natural choice.
The (temporal) data permutation into SoA then is delegated to the
compiler.
For this we introduce C++ attributes
(Algorithm~\ref{algorithm:annotations:example}):

\begin{itemize}
  \item Whenever the compiler encounters a
  \texttt{[[clang::soa\_conversion\_target]]} attribute, we instruct the
  translator to convert the array named as argument into SoA prior to
  entering the following loop. All attribute accesses within the loop
  will be redirected to the converted data. We instruct the compiler to perform
  a temporary out-of-place transformation
  \cite{Sung:2012:DataLayoutTransformations}.
  \item To trigger the conversion, users have to specify the size of the
  target array through \texttt{[[clang::soa\_conversion\_target\_size]]}, as
  plain C arrays do not come along with such meta information.
  \item We can convert the whole struct, i.e.~all attributes, hosted within an
  AoS, or we can restrict the conversion to particular attributes of these
  structs through \texttt{[[clang::soa\_conversion\_inputs]]}. The input annotation opens a
  \emph{view} on the struct. It peels the struct
  \cite{Hundt:2006:StructureLayoutOptimisation}.
  \item Alterations to the implicitly created SoA view become invalid once we
  leave the code block. If changes should be copied back into the original data,
  i.e.~\emph{synchronised}, users have to add
  \texttt{[[clang::soa\_conversion\_outputs]]}.
\end{itemize}

\subsection{Transformation semantics}

Let $S$ be a struct and $\mathbb{S} = [S_0, S_1, S_2, \ldots]$ be a sequence,
i.e.~array of these structs (AoS).
We assume that $\mathbb{S}$ is identified through a pointer to the first element of
the sequence.
As we work with raw data in a C sense, we require explicit information on the
size $|\mathbb{S}|$ of $\mathbb{S}$ from the user,
though passing
\texttt{s.size()}
is admissible if we work with C++'s \texttt{std::array} or
\texttt{std::vector}, e.g.

\texttt{[[clang::soa\_conversion\_target]]} over $\mathbb{S}$ identified
by its pointer informs us that $\mathbb{S}$ is held as AoS, and it introduces a
mapping $aos\_to\_soa: \mathbb{S} \mapsto \mathbb{\hat S}$.
$\mathbb{\hat S}$ is a tuple of data, where each entry is an array of a
primitive type with the size $|\mathbb{S}|$.
With \texttt{[[clang::soa\_conversion\_inputs]]}, the set $\mathbb{A}$ of
attributes that are mapped can be restricted.
Not every attribute of $S$ has to be copied into the reordered data area.

Our compiler does not automatically copy content from the temporary data
structure back.
It does not automatically synchronise $\mathbb{\hat S}$ and $\mathbb{S}$.
Instead, we require users to specify which attribute set $\mathbb{\hat A}$ to
copy back through \linebreak \texttt{[[clang::soa\_conversion\_outputs]]}.
\texttt{[[clang::aos\_conversion\_target]]} introducing $soa\_to\_aos$ and its counterpart \texttt{[[clang::aos\_conversion\_outputs]]} are
well-defined the other way round.

\paragraph{Out-of-place temporary data.}

Let $\mathbb{\hat S}$ hold attributes from $\mathbb{A} \cup \mathbb{\hat A}$.
That is, the temporary data hold all attributes which are either defined through
input or output statements.
We have no subset relation and allow $\mathbb{A} \cap \mathbb{\hat A} \not=
\emptyset$.
With this, $mem(\mathbb{\hat S}) \leq mem(\mathbb{S})$ if $mem$ yields the
memory footprint.
We furthermore note that $soa\_to\_aos = aos\_to\_soa^{-1}$ whenever $\mathbb{A}
= \mathbb{\hat A}$.

\paragraph{Prologue.}

The target statement adds a preamble to the following loop.
The preamble opens with the creation of a temporary data structure
$\mathbb{\hat S}$, and then immediately fills the image with 

\vspace{-0.2cm}
\begin{equation}
  \hat S.a_i \gets S_i.a \qquad \forall 0 \leq i < |\mathbb{S}|, \ \forall a
  \in \mathbb{A}.
  \label{equation:annotation:AoS_to_SoA}
\end{equation}

\noindent
Only attributes $a \in \mathbb{A}$ of $\mathbb{S}$ are copied
into the SoA helper data structure
due to (\ref{equation:annotation:AoS_to_SoA}).
For $\mathbb{A} \not= \mathbb{\hat A}$, we do not fill all attributes in $\mathbb{\hat S}$.
Some fields hold garbage.

\paragraph{Redirection.}

Any subsequent access to $S_i.a, \ a \in \mathbb{A}
\cup \mathbb{\hat A}$ within the loop is redirected to $\hat S.a_i$.
Our compiler extension redirects accesses to the structs within $\mathbb{S}$ to
the structure of arrays.
Any follow-up optimisation pass within the compiler pipeline will hence work
with SoA.
Through an explicit specification of $\mathbb{A}$ and $\mathbb{\hat A}$,
we can shrink the memory footprint of $\mathbb{\hat S}$ and reduce the copy
overhead.
Attribute accesses $a \not \in \mathbb{A}
\cup \mathbb{\hat A}$ continue to hit $\mathbb{S}$
directly.

\paragraph{Epilogue.}

The epilogue adds 

\vspace{-0.2cm}
\begin{equation}
  S_i.a \gets \hat S.a_i \qquad \forall 0 \leq i < |\mathbb{S}|, \ \forall a
  \in \mathbb{\hat A}.
  \label{equation:annotation:SoA_to_AoS}
\end{equation}

\noindent
to the outcome code, followed by a free of $\mathbb{\hat S}$.
In principle, the epilogue uses
$soa\_to\_aos$.
The difference between (\ref{equation:annotation:SoA_to_AoS}) and
(\ref{equation:annotation:AoS_to_SoA}) is the temporal order: 
An output statement inserts the mapping into an epilogue, while the target
statement tackles the prologue.

\subsection{Extensions}

Our language extension is focused on one-dimensional data sets which are
logically continuous, i.e.~allow for coalescent data access: 
A container traversed via \texttt{for (auto\& p: ...)} can technically be
realised with scattered data $\mathbb{S}$.
Our conversion yields one continuous excerpt $\mathbb{\hat S}$.

We implicitly exploit that any loop imposes an ordering.
Therefore, we can extend our annotations by 
\texttt{[[clang::soa\_conversion\_start\_idx()]]} 
which restrict the conversion to a subset of the range.
There is no need for a \linebreak
\texttt{[[clang::soa\_conversion\_end\_idx()]]},
as the size is manually specified.

\section{Compiler realisation}
\label{section:realisation}

\begin{algorithm}[htb]
  \begin{algorithmic}[1]
 \Function{void drift}{Particle *particles,  int size}
 \State pos0$_{\text{tmp}}$ = new double[size];
   \Comment{temporary out-of-place array}
 \State pos1$_{\text{tmp}}$ = new double[size];
 \State \ldots
 \For{(int i = 0; i $<$ size; i++)}
   \State pos0$_{\text{tmp}}$[i] = particles[i].pos[0];
     \Comment{out-of-place AoS-SoA conversion}
   \State \ldots
 \EndFor
 \For{(int i = 0; i $<$ size; i++)}
   \State pos0$_{\text{tmp}}$[i] += vel0$_{\text{tmp}}$ * dt;
     \Comment{p.pos[0] += p.vel[0] * dt;}
   \State \ldots
   \State updated$_{\text{tmp}}$[i] = true;
     \Comment{p.updated = true;}
 \EndFor
 \For{(int i = 0; i $<$ size; i++)}
   \State particles[i].pos[0] = pos0$_{\text{tmp}}$[i];
     \Comment{SoA-AoS conversion due}
   \State \ldots
     \Comment{to outputs statement}
 \EndFor
 \State delete[] pos0$_{\text{tmp}}$;
   \Comment{free temporary arrays}
 \State \ldots  
 \EndFunction
\end{algorithmic}

  \caption{
    Pseudo code which illustrates how the compiler rewrites the example in
    Algorithm~\ref{algorithm:annotations:example}.
    Embedded arrays where the size has to be known at compile time and types of
    all converted data can be analysed by the compiler from the source code.
    \label{algorithm:annotations:example-converted}
    }
\end{algorithm}

We implement our proposed techniques prototypically such that we can highlight
their potential impact for the SPH demonstrator.
For this, we plug into the Clang LLVM front-end.
Clang takes C++ code and emits unoptimised LLVM IR.
As we work within Clang, our modifications affect the resultant unoptimised intermediate representation (IR) output, and do not alter any downstream steps within the
translation pipeline. Notably, the generated IR code still benefits from all LLVM-internal optimisation passes.
Clang's lexer, parser and semantic analyser yield an
abstract syntax tree (AST).
The AST then is subsequently consumed to produce LLVM IR.
These steps constitute the compiler's \texttt{FrontendAction}.

We realise our functionality with a new \texttt{FrontendAction}.
It traverses the AST top down and searches for our 
annotations.
When it encounters a convert, it
inserts allocation statements for the out-of-place memory allocations
once all information on the data view are available, and it adds
the actual data copying from
(\ref{equation:annotation:AoS_to_SoA}).
The names of the temporary variables are mangled to avoid variable
re-declarations.
Following the prologue step, we query the actual loop body and redirect the
memory accesses to $\mathbb{A} \cup \mathbb{\hat A}$.
Finally, we add an epilogue that synchronises the data according to
(\ref{equation:annotation:SoA_to_AoS}) and frees the temporary buffers.
As AST modifications are not possible in Clang, i.e.~as the AST is
immutable, 
our realisation pretty prints the AST into C++
subject to our modifications and then invokes Clang's
original \texttt{FrontendAction}.
This process happens in-memory, and is transparent to the end user.
The behaviour can be disabled on the command line through 
\texttt{-fno-soa-conversion-attributes-language-extension}.
Our compiler optimisation pass materialises in a source-to-source compiler
(cmp.~rewrite of Algorithm~\ref{algorithm:annotations:example}  into
\ref{algorithm:annotations:example-converted}) which feeds into subsequent
translation sweeps \cite{Vikram:2014:LLVM}.

As we precede Clang's original \texttt{FrontendAction}, it is possible to dump
our output into source files explicitly instead of implicitly passing it on into
the subsequent \texttt{FrontendAction}.
While originally designed for debugging, this feature is particularly appealing
in environments where the modified compiler is only available locally,
while the compilers on the target production platform cannot be
modified.

As we navigate the source code using the AST, the effect of our annotations is scoped at the translation-unit level:
The data transformations cannot propagate into user translation units (object
files) and notably fail to propagate into libraries.
Instead, the annotations will break the code if library functions are used
internally.
While functions with signatures similar to
\texttt{foo(particles[i])} (cmp.~Algorithm~\ref{algorithm:annotations:example}) cannot be
used in code blocks subject to the code transformations, functions capturing
attributes individually are supported.
A function \texttt{foo(particles[i].pos[0],}
\texttt{particles[i].pos[1],\ldots)} will continue to work.
A sequence of \texttt{foo} calls will even vectorise
if translated accordingly (\texttt{declare simd} in OpenMP for example).

\section{Benchmark results}
\label{section:results}

We assessed the impact of our compiler prototype on two architectures.
Our first system is an AMD EPYC 9654 (Genoa) testbed. 
It features $2 \times 96$ cores over $2 \times 4$ NUMA domains spread over
two sockets, hosts
an L2 cache of 1,024 KByte per core and a shared L3 cache with 384 MByte per
socket.
%
%
%
Our second system is an Intel Xeon Gold 6430 (Sapphire Rapid).
It features $2 \times 32$ cores over two sockets.
They form two NUMA domains with 
an L2 cache of 2,048 KByte per core and a shared L3 cache with 62 MByte per
socket.
\added[id=R03]{
To avoid NUMA effects, we confine all our Intel and AMD tests to a single NUMA
domain through \texttt{numactl}.
}

\begin{figure}[htb]
  \begin{center}
    \includegraphics[width=0.44\textwidth]{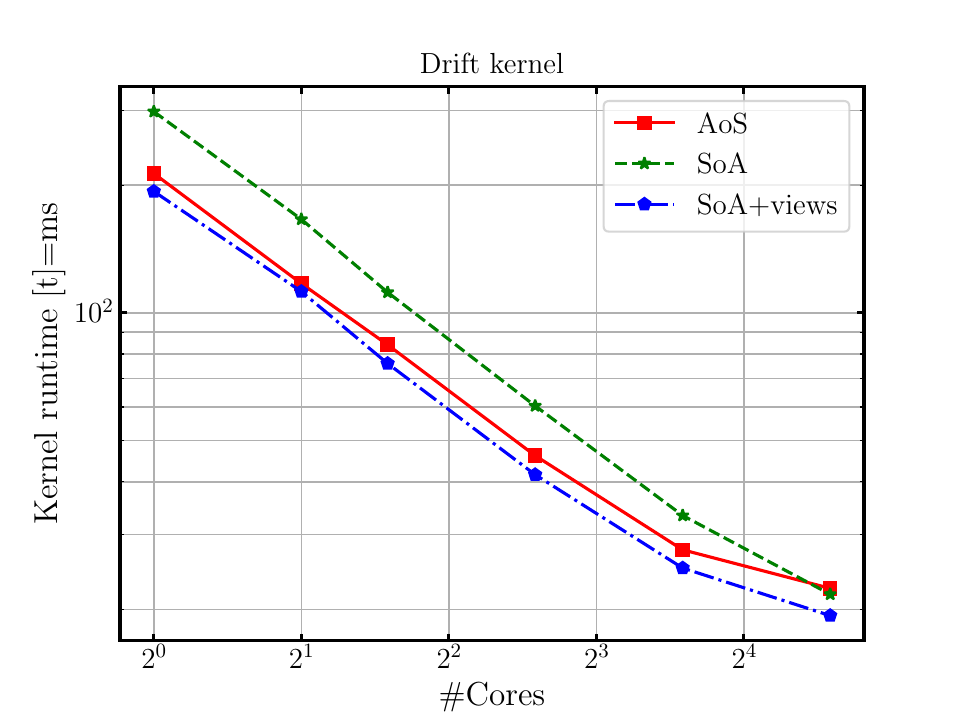}
    \includegraphics[width=0.44\textwidth]{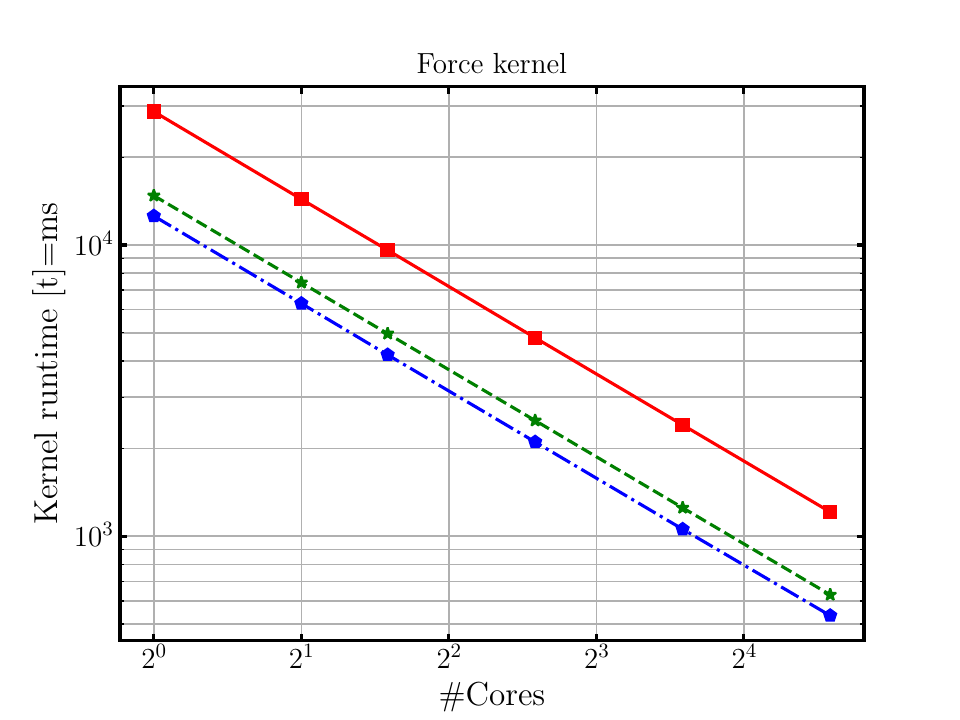}
    \includegraphics[width=0.44\textwidth]{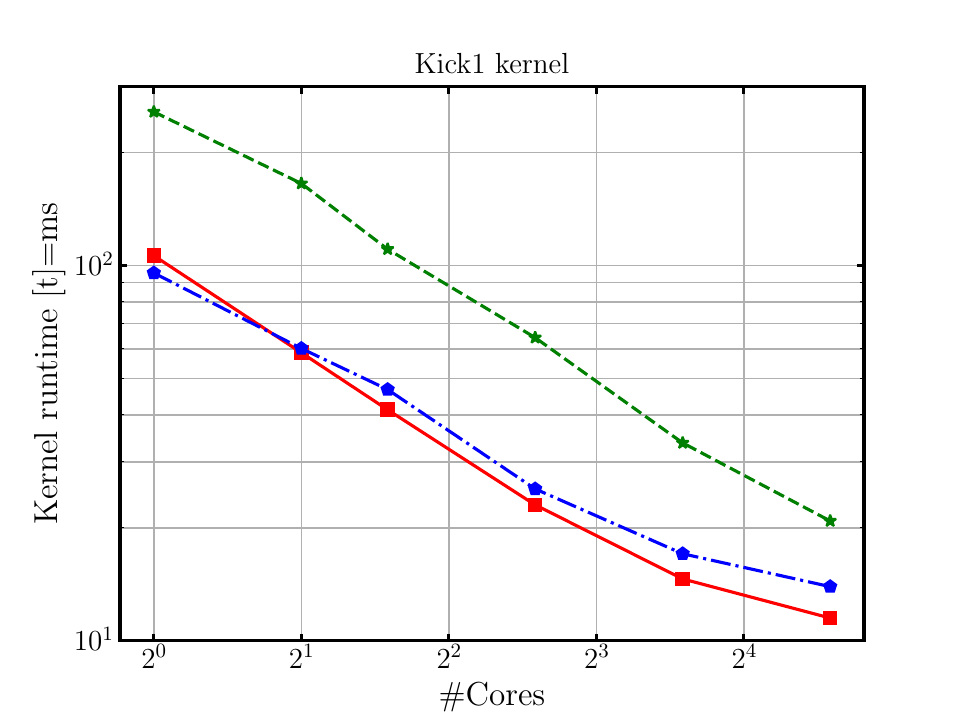}
    \includegraphics[width=0.44\textwidth]{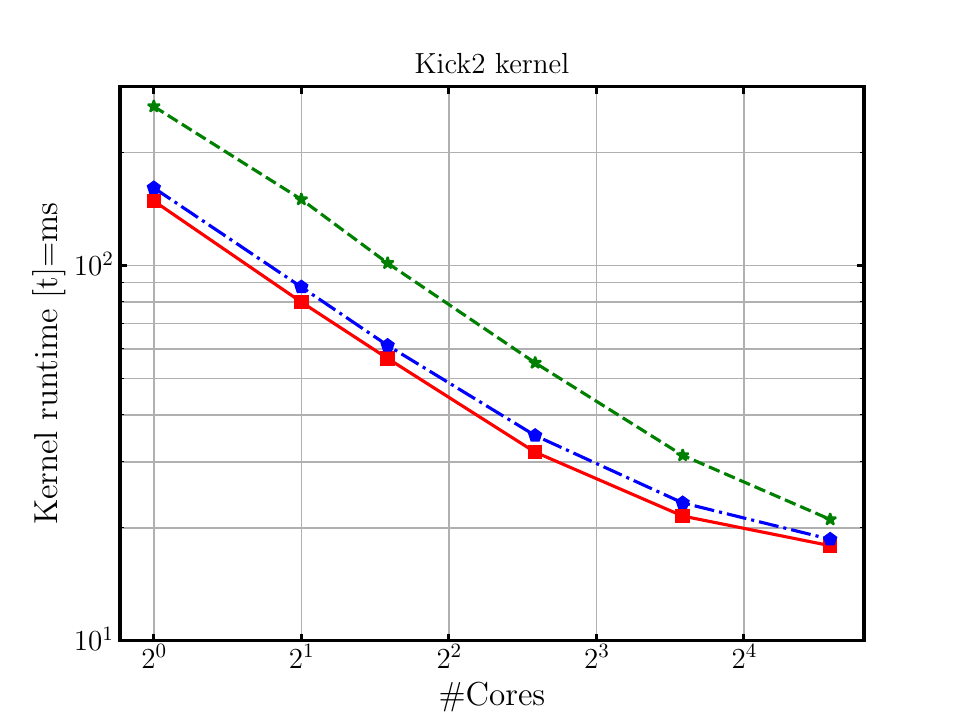}
  \end{center}
  \vspace{-0.6cm}
  \caption{
    Scalability plots for $2048^2$ particles on a single node for four
    different kernels (Genoa). We benchmark the baseline code (AoS)
    against a version which converts all of the particle data (SoA) against a version which works with
    views.
    \label{figure:results:scaling:genoa}
  }
\end{figure}

\paragraph{Upscaling.}
%
%
We start our studies with a classic strong scaling configuration for
$4.19 \cdot 10^6$ particles.
The particles are held as AoS and we iterate over them with a parallel loop
invoking one of our kernels only.
Our measurements distinguish three different variants:
A plain AoS serves as vanilla version. 
This is the code base actually written down in C++.
We compare this to SoA where the whole particle sequence is converted prior to
loop and synchronised back afterwards, i.e.~$\mathbb{A} = \mathbb{\hat A}$ with all attributes of the struct
involved.
Finally, we assess a version where views narrow down the attribute sets to the
number of variables which are actually read or written, respectively.

%
%
The force calculation scales almost perfectly\added[id=R03]{ as it is
compute-bound}, while kicks and drifts tail off
(Figure~\ref{figure:results:scaling:genoa}).
This implies that the latter ones become bandwidth-bound once we use a
sufficiently high number of cores. 
Converting into a SoA leads to a higher compute time, but
narrowing down the conversion to views eliminates this penalty.
For the drift, the converted version becomes even slightly faster than the 
AoS vanilla version.
The situation is fundamentally different for the computationally demanding force
kernel.
Here, the temporary conversion into SoA pays off, and the views then help to
reduce the runtime even further.
All measurements for the Sapphire Rapid yield qualitatively similar data.

%
%
Adding additional data movements due to AoS--SoA
conversions to computationally cheap kernels is poisonous: 
Even an improved AVX512 vectorisation cannot compensate for the logical latency
that our out-of-place conversion introduces. 
We delay the actual loop launch as we first have to convert data.
If we employ many cores and operate
overall in the saturation regime, the difference starts to disappear.
As a core has to wait for the memory controller anyway, it can as well use
the cycles to convert the data structure in a nearby cache.

If a kernel exhibits sufficient compute load, a temporary conversion pays off.
This contradicts the intuition that a conversion is particularly beneficial for
Stream-like operations
\cite{Intel:MemoryLayoutTransformations,Homann:2018:SoAx}.
It aligns with other papers where temporary transformations are used for
expensive code parts \cite{Gallard:2020:ExaHyPEVectorisation}.

\paragraph{Impact of particle count.}
%
%
We continue with a test where we alter the number of particles per kernel, 
draw 16 samples per run to avoid noise, and study the throughput, i.e.~the
number of particle updates or the number of particle--particle interactions
per second, respectively.
As the scalability of the benchmark is confirmed, we stick to three
characteristic thread counts.
As we have demonstrated the impact of the views, we focus on a comparison of 
the views to the vanilla AoS storage.

\begin{figure}[htb]
  \begin{center}
    \includegraphics[width=0.44\textwidth]{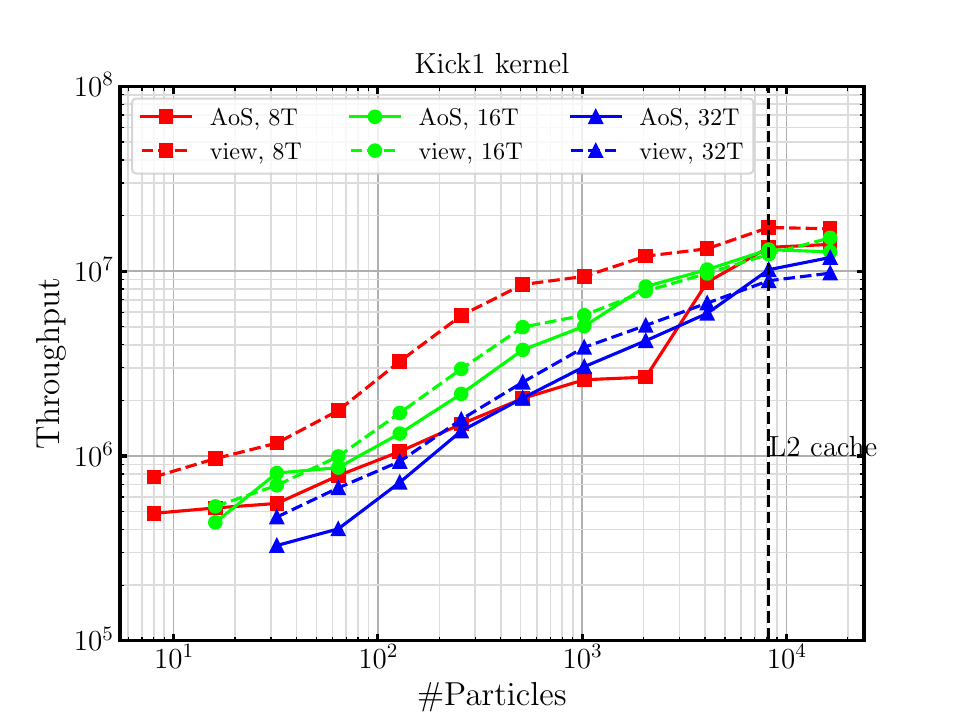}
    \includegraphics[width=0.44\textwidth]{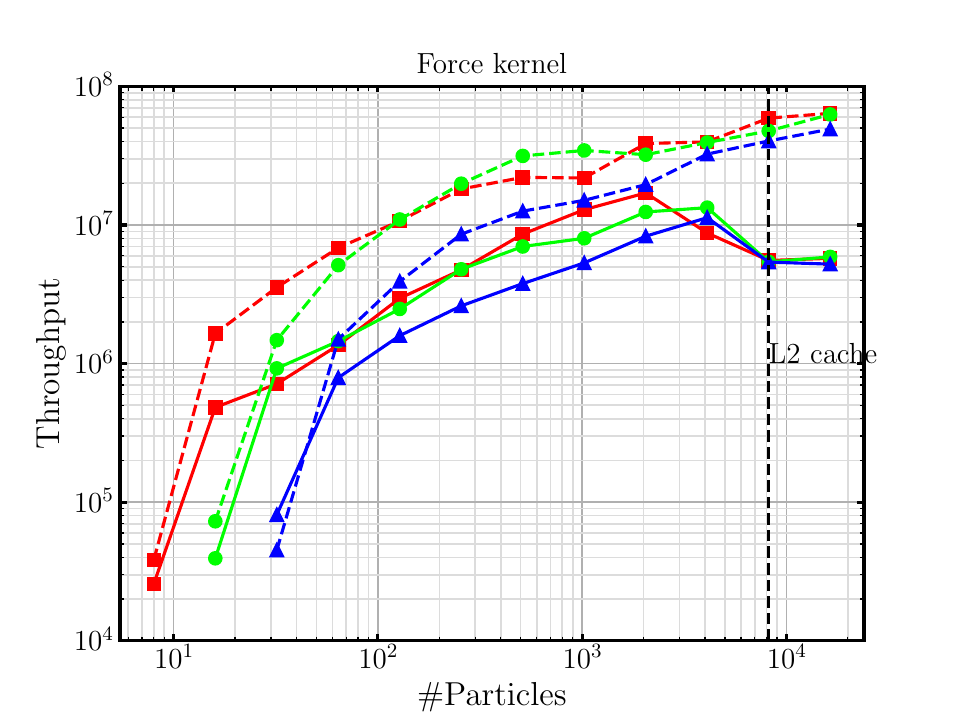}
    \\
    \includegraphics[width=0.44\textwidth]{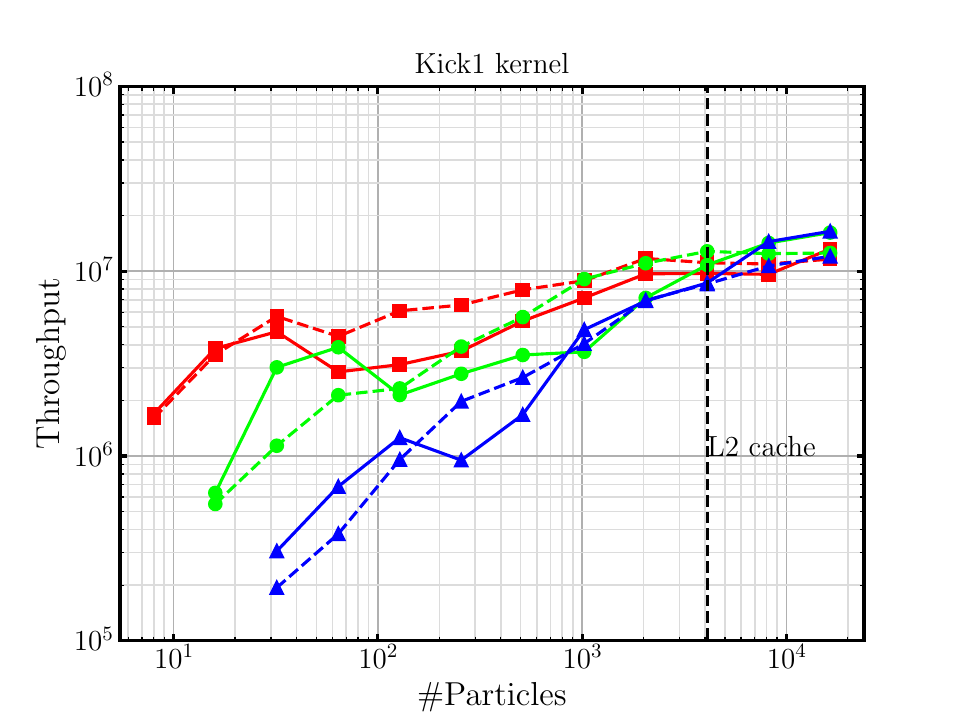}
    \includegraphics[width=0.44\textwidth]{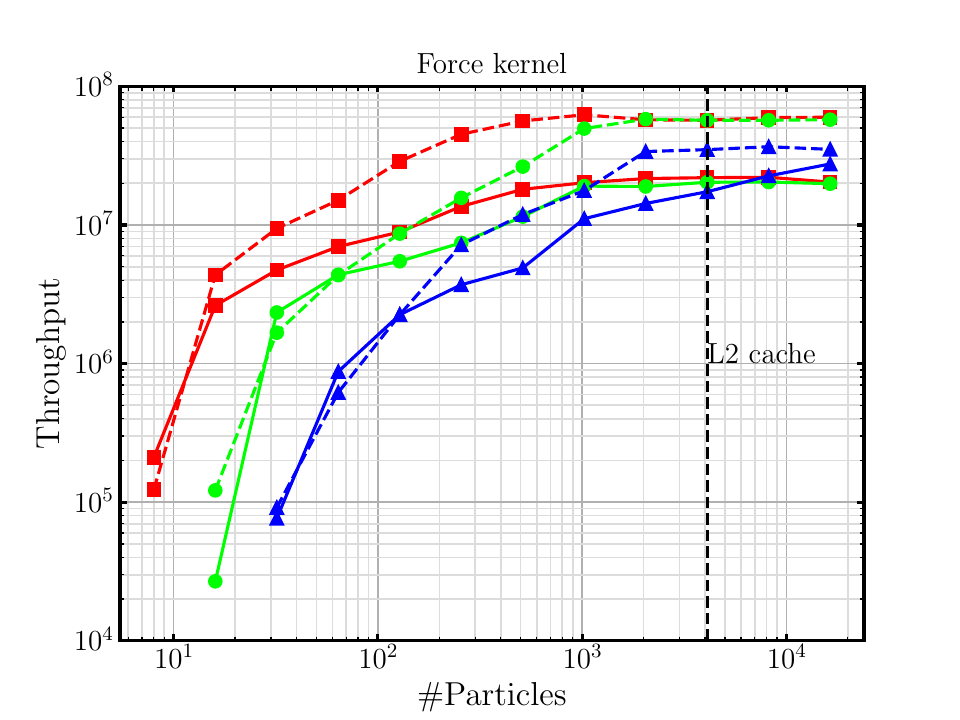}
  \end{center}
  \vspace{-0.6cm}
  \caption{
    \noindent
    Dependency on particle counts for
    fixed thread numbers (8, 16 and 32). Sapphire Rapid (top) vs.~Genoa data
    (bottom).
    The L2 cache size denotes the size for a single thread, i.e.~how many
    particles would fit into one single L2 cache.
    \label{figure:results:throughput:sapphirerapid}
  }
\end{figure}

%
%
For the cheap kernels, the conversion pays off as long as the particle count is
\replaced[id=R04]{contained within a machine-specific optimal range}{sufficiently high yet not to big}.
The efficiency gap closes on Intel systems as we use more threads, and becomes
rather erratic on AMD (Figure~\ref{figure:results:throughput:sapphirerapid}).
``\replaced[id=R04]{Optimal range}{Not to big}'' notably implies that we are not in a scaling regime yet, where
using all threads pays off.
For the compute-heavy force kernel, the gap between sole AoS and the views
widens on the Intel system as the particle count increases.
We notably observe a ``sudden'' throughput deterioration for the plain AoS which
does not occur for the converted algorithm.
On AMD, the curves for both schemes are relatively smooth, yet they close again
if we use too many threads:
The SoA version with views stagnates, while the AoS version catches up.

%
%
The large caches on both testbeds imply that we
never fall out of L3 cache once the first warm-up run out of 16 is
completed.
We hence rule out last level cache misses to explain the throughput behaviour.
Our L3 is filled with the AoS data once per test, and these data are
then converted 16 times into SoA with views and back.
We manually verified that the machine code (AVX vectorisation) is comparable on
both target systems and is efficient.
Runtime behaviour and qualitative differences in the data thus have to
be explained through the memory access characteristics.

We need a decent number of particles per thread to give the vectorisation the
opportunity to unfold its potential.
For very small particle counts per thread, a conversion never pays off.
For small particle counts per thread, we however benefit within the
loop body from improved memory access characteristics.
Once the particle count increases, the advantage is eaten up by the
single-threaded conversion in the epilogue and preamble of the loop if we encounter an algorithm with linear complexity and low
computational load.
If we have ``too many'' threads, they have to synchronise through the shared L3
cache, which adds a further penalty and renders the conversion useless.

For high computational load and quadratic complexity, the improvement in better
memory access characteristics outweighs the conversion penalty.
The performance gap widens.
On the Intel system, the original AoS version takes a hit once we fall out of
the L2 cache on a core.
The converted SoA memory comprises only the hot data~\cite{Vikram:2014:LLVM},
i.e.~the data we actually work on.
It continues to fit into the L2. 
On the AMD system, we however quickly start to suffer from NUMA effects once we
employ too many threads.
We have to give up on our performance wins.

\paragraph{Context.}
Our demonstrator uses structs that are embedded into other
structs~\cite{Springer:2018:SoALayout}, as we work with coordinate vectors
within the particle structure.
It supports data hierarchy through struct composition, but seems to restrict
itself exclusively to AoS and SoA.
It is important to realise however that we actually support more sophisticated
data structures, too.
This observation results directly from the fact that our conversion is local and
temporary.

Our underlying code base \cite{Schaller:2024:Swift} and many other codes
advocate for the use of AoSoA \cite{Gallard:2020:Roles} or ASTA
\cite{Sung:2012:DataLayoutTransformations} to compromise between speed and
flexibility.
As we may convert subarrays of a given AoS structure, our approach can logically
yield an AoSoA/AoS hybrid.
The overall data are stored as AoS, but subsections flip to SoA.
Along the same lines, our techniques allow for the manipulation of data
structures where chunks of AoS data are scattered over the heap.
It does not require one global data space.
Both properties are important for our SPH code base, where cells for example
host a pointer.
The data per cell are held as AoS, but the cells' data are scattered
over the heap.


As long as SPH codes run through the individual algorithmic phases step by step
and synchronise after each step in a fork-join manner, only the algorithmic phases with quadratic complexity per
control volume seem to benefit.
However, once we work with codes that work with functional decomposition
(task-based parallelism) or employ multiple ranks per node, simulations tend
to run different algorithmic steps on different threads concurrently.
As as result, fewer threads than the whole node configuration are effectively 
available to a compute step.
It might be reasonable to parallelise the conversion into SoA and
back.
In practice, it is not clear if this adds significant value as a step rarely
``owns'' many or all threads.


\section{Conclusion and outlook}
\label{section:conclusion}

%
%
We introduce a local, guided approach to convert from AoS into SoA and
vice versa within simulation codes\replaced[id=R04]{.}{:} 
When users add our novel C++ annotations to their code, the compiler rewrites
its underlying data structure out-of-place into a temporary scratchpad for the
affected code block, redirects all source code operations to this altered memory
location, and eventually synchronises (parts of) the temporary data structure
with its original again.
Our approach allows for the incremental, non-disruptive optimisation of code, as
all code remains valid if a compiler does not support our annotations, and as
users can continue to design their implementation with their favourite data
layout in mind \cite{Gallard:2020:Roles}.
In object-oriented codes, this will likely be AoS.
Our experimental studies challenge the wide-spread assumption that such
temporary, local reordering is not affordable.
It is affordable once we introduce the notion of a view, i.e.~the opportunity to
restrict transformations only to some attributes of a struct.

%
%
Our data are obtained through a prototypical Clang compiler
extension.
A more mature, yet more heavy-weight long-term realisation might implement the
transformations within LLVM's MLIR layer and hence make it independent of the
used front-end.
The promise here is that the conversion would automatically benefit 
from other memory optimisations such as the automatic introduction of 
padding and alignment. 
Within the C++ domain, it is an interesting challenge for future work to 
discuss how the data transformations could be integrated with C++
views or memory abstractions as pioneered within Kokkos
\cite{Trott:2022:Kokkos}, as the conversion gain is clearly tied to the
underlying problem sizes and data layouts.
Both directions of travel for the compiler construction are timely and
reasonable, as our work has demonstrated that temporal out-of-place conversions have the potential to speed
up code. 
In particular, it is reasonable to assume that our idea unfolds its full
potential once we apply it to
GPU-based codes~\cite{Sung:2012:DataLayoutTransformations}.
\added[id=R02]{
 It also feeds into a high-level discussion whether data transformations should
 be offered to developers \cite{Childers:2024:CPPReflections} or realised by and
 within the compiler subject to developer guidance.
}

%
%
There is an elephant in the room: 
We leave the responsibility to insert appropriate data reordering instructions
with the user and do not try to move the decisions themselves into the compiler
\cite{Hundt:2006:StructureLayoutOptimisation}.
Designing an automatic data transformation requires
appropriate heuristics.
Our studies suggest that such a heuristic is inherently difficult to construct,
as it would require a ``what if'' analysis:
what would the gain of the transformation be, i.e.~how much performance
improvement would unfold due to better vectorisation, reduced cache misses and
reduced TLB, and could those improvements compensate for any transformation
overhead.
Each of these questions has to be answered over a huge space of potential
reorderings.
We hence hypothesise that only feedback optimisation (static) or on-the-fly
adjourning and just-in-time compilation can deliver
beneficial automatic reordering
\cite{Hundt:2006:StructureLayoutOptimisation,Vikram:2014:LLVM}.

A second train of thought arises from the insight that conversions benefit
greatly from the concept of views, i.e.~partial permutations and copying of the
data structures.
Future work has to study how we can weaken the
notion of temporal, localised
transformations~\cite{Hirzel:2007:DataLayoutForOO} further:
It is a natural idea to preserve converted data over multiple code blocks,
i.e.~not to free and re-allocate them but to synchronise selectively.
This would lead to a setup where data is held both in \replaced[id=R1]{AoS}{SoA}
and SoA---and both subject of different views---and the compiler automatically synchronises these
representations upon demand.
We expect this to reduce overhead massively, trading memory for speed.

\section*{Acknowledgments}

Tobias' research has been supported by EPSRC's Excalibur
programme through its cross-cutting project EX20-9 \textit{Exposing Parallelism: Task Parallelism}
(Grant ESA 10 CDEL) and the DDWG projects \textit{PAX--HPC} (Gant EP/W026775/1)
as well as \textit{An ExCALIBUR Multigrid Solver Toolbox for ExaHyPE}
(EP/X019497/1).
His group appreciates the support by Intel's Academic Centre of
Excellence at Durham University. 
The work has received funding through the eCSE project ARCHER2-eCSE11-2
``ExaHyPE-DSL''.

The code development relied on experimental test nodes installed within  
the DiRAC@Durham facility managed by the Institute for Computational Cosmology
on behalf of the STFC DiRAC HPC Facility
(\href{www.dirac.ac.uk}{www.dirac.ac.uk}). The equipment was funded by BEIS capital funding via STFC capital grants ST/K00042X/1, ST/P002293/1, ST/R002371/1 and ST/S002502/1, Durham University and STFC operations grant ST/R000832/1. DiRAC is part of the National e-Infrastructure.

\bibliographystyle{splncs04}
\bibliography{paper}

\newpage

\appendix

\section{Compiler download}
\label{appendix:compiler-download}

The forked Clang/LLVM project is publicly available at
\linebreak 
\texttt{https://github.com/pradt2/llvm-project}, our extensions are available in the \texttt{hpc-ext} branch.
To avoid conflicts, it is strongly recommended to remove any existing Clang/LLVM installations before proceeding.

Once cloned, we create a build folder in the cloned repository, and
afterwards execute the following command:

\begin{verbatim}
cmake 
-DCMAKE_BUILD_TYPE="RelWithDebInfo" 
-DCMAKE_C_COMPILER="gcc" 
-DCMAKE_CXX_COMPILER="g++" 
-DLLVM_ENABLE_PROJECTS="clang;openmp" 
-DLLVM_TARGETS_TO_BUILD="host" 
-DBUILD_SHARED_LIBS="ON" 
-G "Unix Makefiles" 
../llvm
\end{verbatim}

\noindent
\texttt{make -j\$(nproc) \&\& make install} compile and install the compiler.
From hereon, \texttt{clang++} directs to our modified compiler version.


Its command-line interface (CLI) is backwards compatible with the mainline
Clang/LLVM version. 
The support for the new attributes discussed in this paper is enabled by default, no additional compiler flags are needed.
Yet, the features can be switched off (Section~\ref{section:realisation}).

If the use of any of the new attributes leads to a compilation error, a common
troubleshooting starting point is to inspect the rewritten source code. To see
the rewritten code, we can add \texttt{-fpostprocessing-output-dump} to the
compilation flags. This flag causes the post-processed source code be written to the standard output.

\section{Download, build and run testbench}
\label{appendix:download-and-compilation}

All of our code is hosted in a public git repository 
at \url{https://gitlab.lrz.de/hpcsoftware/Peano}. 
Our benchmark scripts are merged into the repository's
main, i.e.~all results can be reproduced with the main branch. 
Yet, to use the exact same code version as used for this paper, please switch to 
the \texttt{particles-aos-vs-soa} branch.

\begin{algorithm}[htb]
    \begin{algorithmic}[1]
      \State git clone https://gitlab.lrz.de/hpcsoftware/Peano
      \State cd Peano
      \State libtoolize; aclocal; autoconf; autoheader; cp src/config.h.in .
      \State automake {-}{-}add-missing
      \State ./configure CXX=... CC=... CXXFLAGS=... LDFLAGS=... \\
        --enable-loadbalancing --enable-particles --enable-swift \\ 
        --with-multithreading=omp
      \State make  
    \end{algorithmic}
  \caption{
    Cloning the repository and setting up the autotools environment. 
    \label{algorithm:appendix:download-and-compile:autotools}
    }
\end{algorithm}

The test benchmarks in the present paper are shipped as part of Peano's
benchmark suite, which implies that Peano has to be configured and built first.
The code base provides CMake and autotools (Alg.~\ref{algorithm:appendix:download-and-compile:autotools})
bindings.
Depending on the target platform, different compiler options have to be used.
Once configured, the build (\texttt{make}) yields a set of static libraries
providing the back-end of our benchmarks.

The actual benchmark can be found in Peano's subdirectory \linebreak
\texttt{benchmarks/swift2/hydro/aos-vs-soa-kernel-benchmarks}.
Within this directory, a sequence of Python commands (Alg.~\ref{algorithm:appendix:download-and-compile:build-benchmark}) produces
the actual kernel benchmark executable \texttt{kernel-benchmarks}.
The Python script provides further options available through the argument
\texttt{--help}.
Internally, it parses the configuration used to build the static library,
creates all glue code required by the benchmark, and then yields a plain 
Makefile.
By default, this Makefile is immediately invoked.

\begin{algorithm}[htb]
    \begin{algorithmic}[1]
      \State cd benchmarks/swift2/hydro/aos-vs-soa-kernel-benchmarks
      \State export PYTHONPATH=../../../../python
      \State python3 kernel-benchmark.py -d 2
    \end{algorithmic}
  \caption{
    Building the benchmark itself. 
    \label{algorithm:appendix:download-and-compile:build-benchmark}
    }
\end{algorithm}

With the executable at hand, we can run the benchmark and obtain an output
similar to

{\tiny
\begin{verbatim}
===============================
  4096 particles (16 samples)
===============================
density kernel: 0.802495  (avg=0.802495,#measurements=16,max=1.52808(value #12),min=0.705117(value #4),+90.4159%,-12.1344%,...
force kernel:   0.640393  (avg=0.640393,#measurements=16,max=1.17305(value #13),min=0.576625(value #5),+83.1761%,-9.95752%,...
kick1 kernel:   4.29028e-05  (avg=4.29028e-05,#measurements=16,max=0.000144243(value #12),min=3.4756e-05(value #4),+236.209%,...
kick2 kernel:   7.64441e-05  (avg=7.64441e-05,#measurements=16,max=0.000265726(value #12),min=6.2503e-05(value #4),+247.608%,...
drift kernel:   1.86415e-05  (avg=1.86415e-05,#measurements=16,max=4.9618e-05(value #12),min=1.632e-05(value #4),+166.17%,...
\end{verbatim}
}

\noindent
which we can postprocess further.

\section{Additional scalability tests}
\label{appendix:scalability-data}

   \begin{center}
    \includegraphics[width=0.4\textwidth]{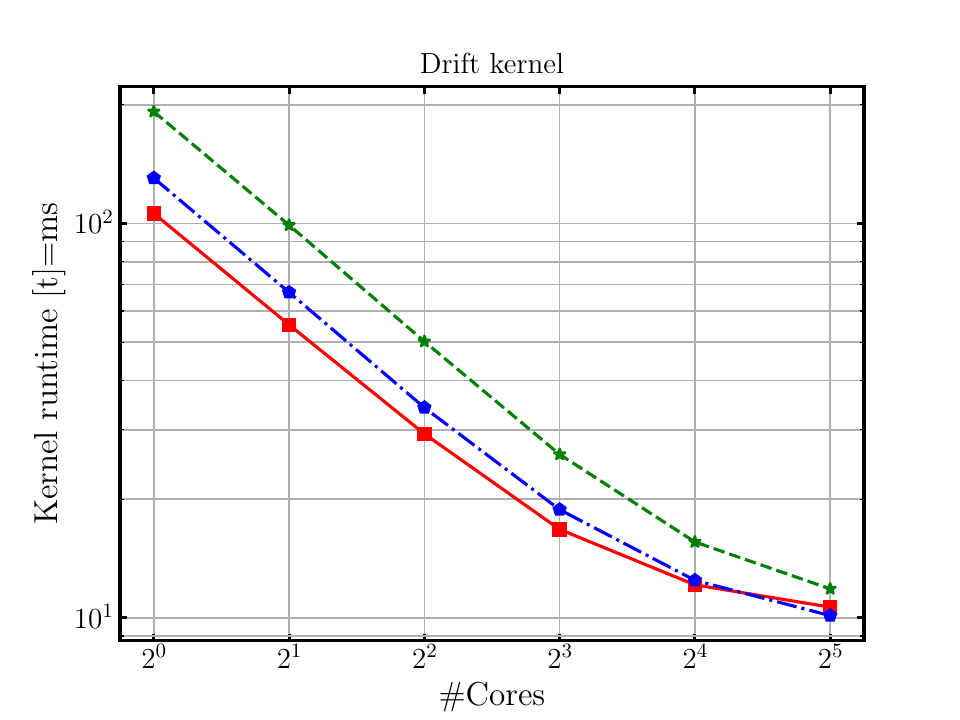}
    \includegraphics[width=0.4\textwidth]{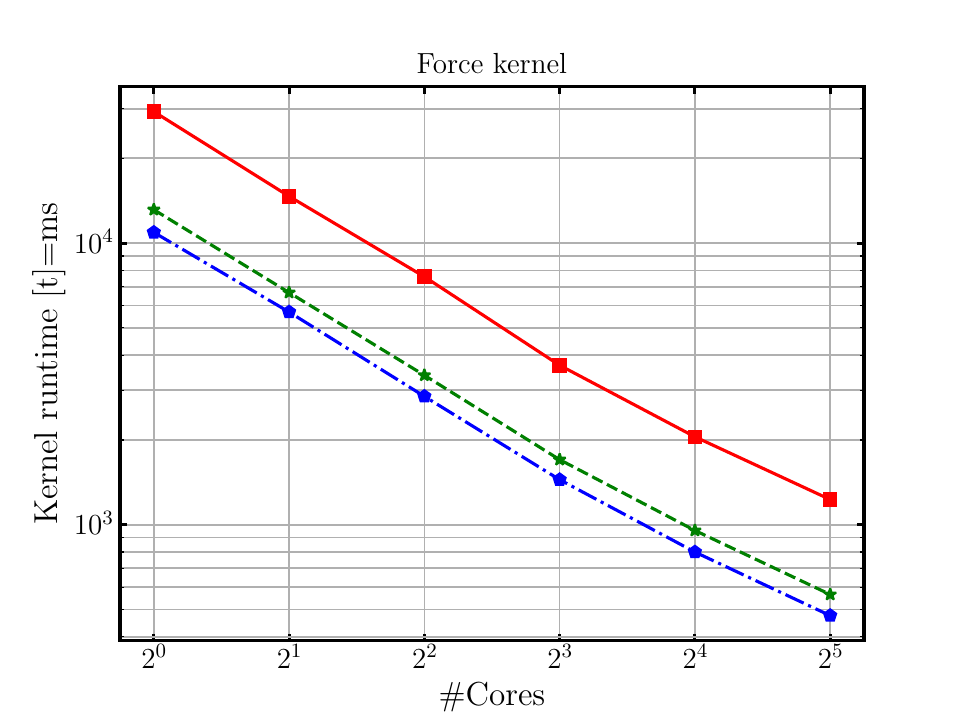}
    \\
    \includegraphics[width=0.4\textwidth]{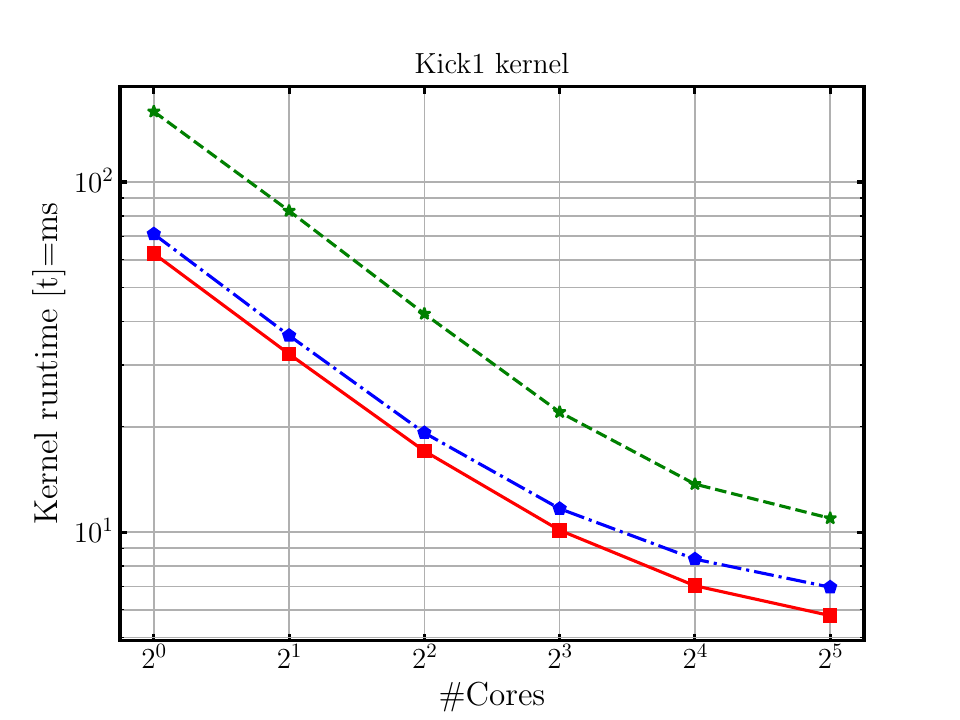}
    \includegraphics[width=0.4\textwidth]{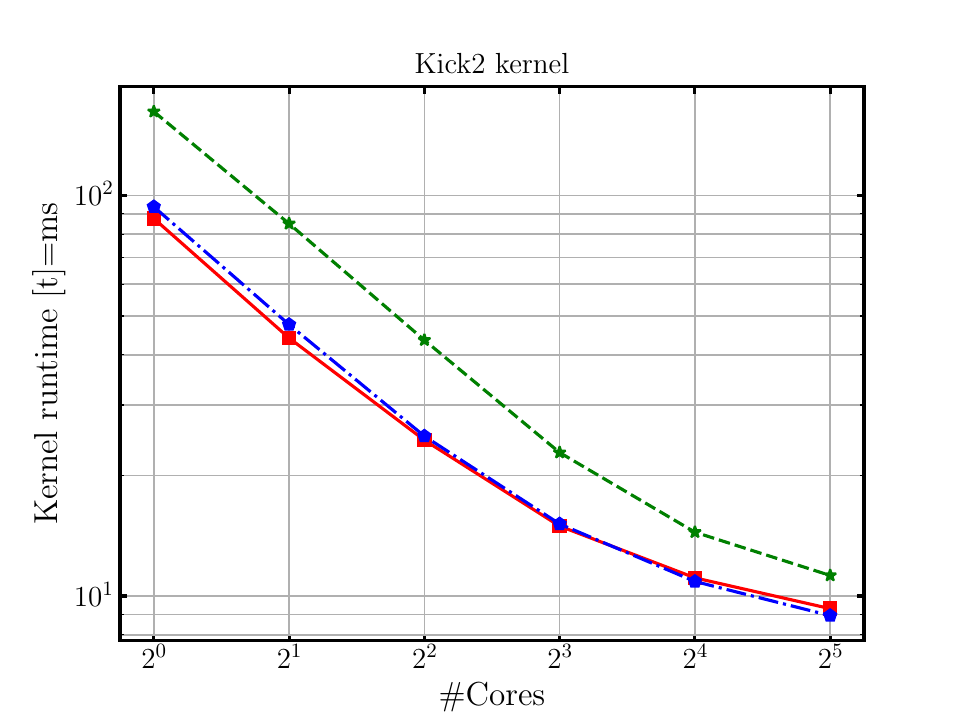}
   \end{center}
    \noindent
    Scalability plots for $2048^2$ particles, i.e.~the same experiment as in
    Figure~\ref{figure:results:scaling:genoa} on the Sapphire Rapid testbed.

  \begin{center}
    \includegraphics[width=0.4\textwidth]{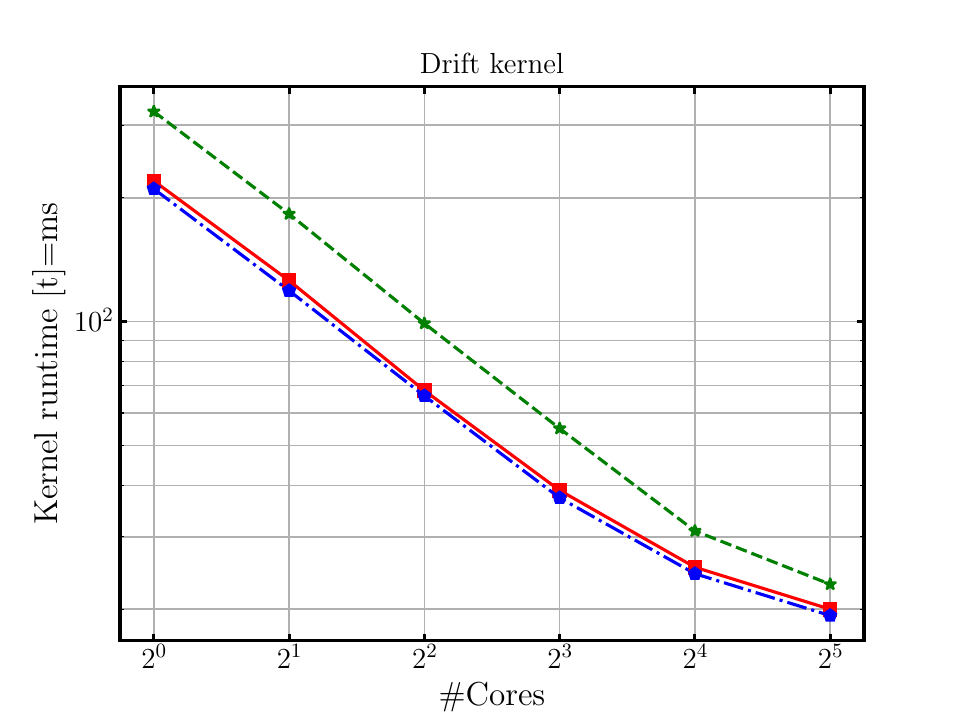}
    \includegraphics[width=0.4\textwidth]{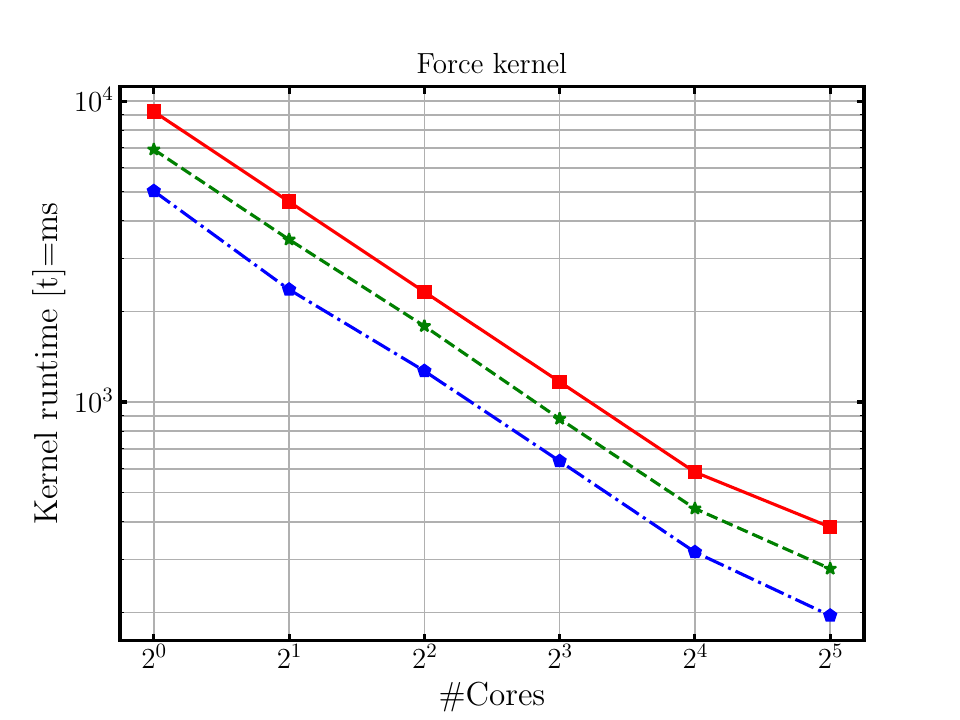}
    \\
    \includegraphics[width=0.4\textwidth]{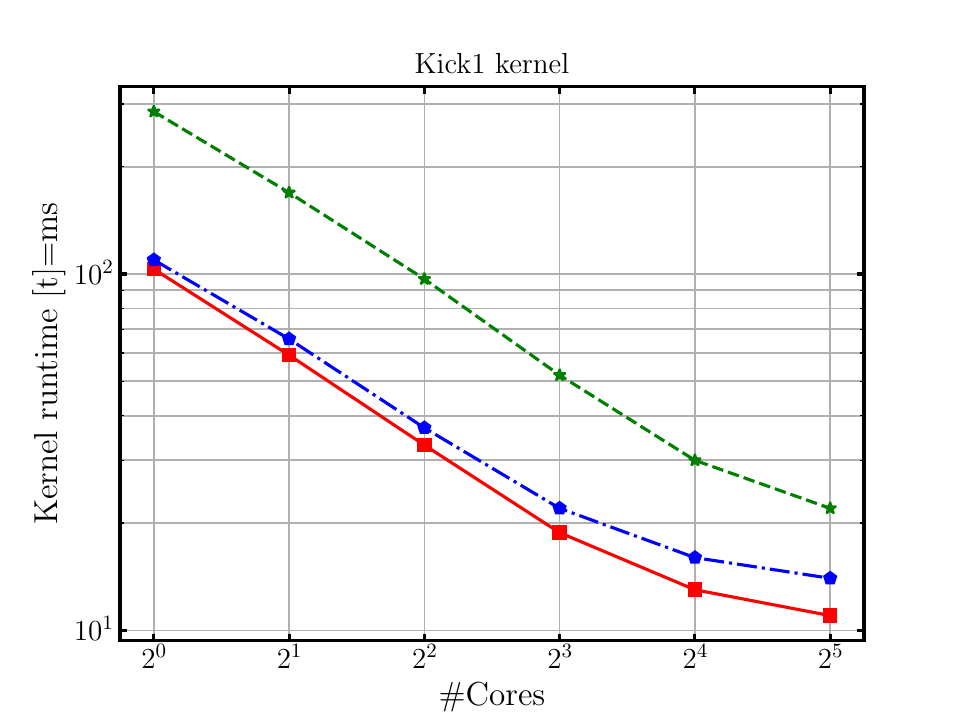}
    \includegraphics[width=0.4\textwidth]{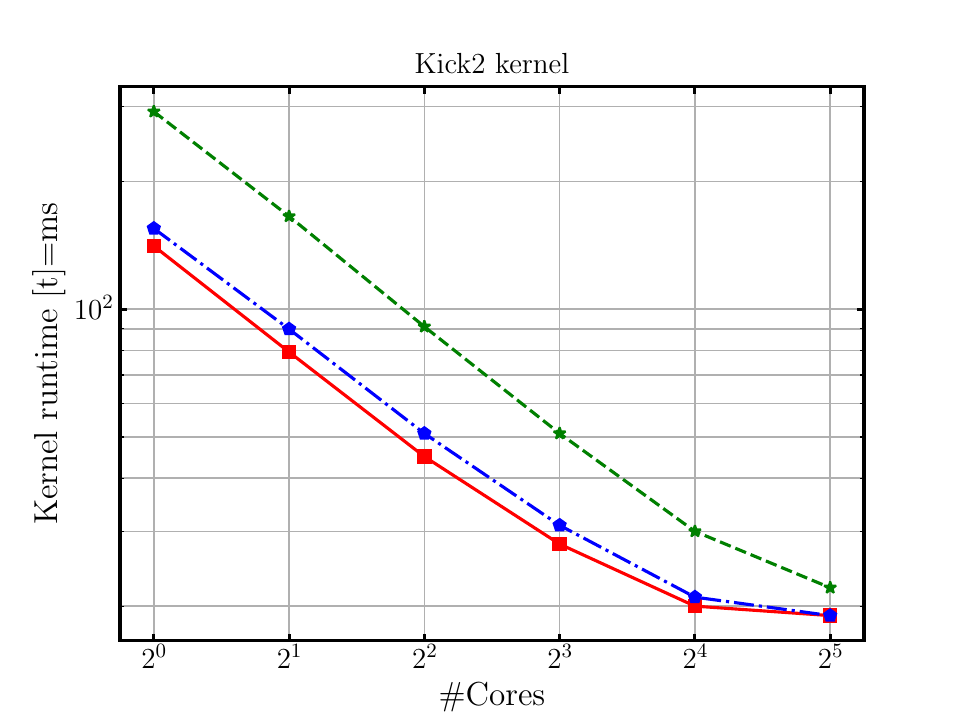}
  \end{center}
    \noindent
    Experiment from Figure~\ref{figure:results:scaling:genoa} for
    $4 \cdot 2048^2$ particles.

  \begin{center}
    \includegraphics[width=0.4\textwidth]{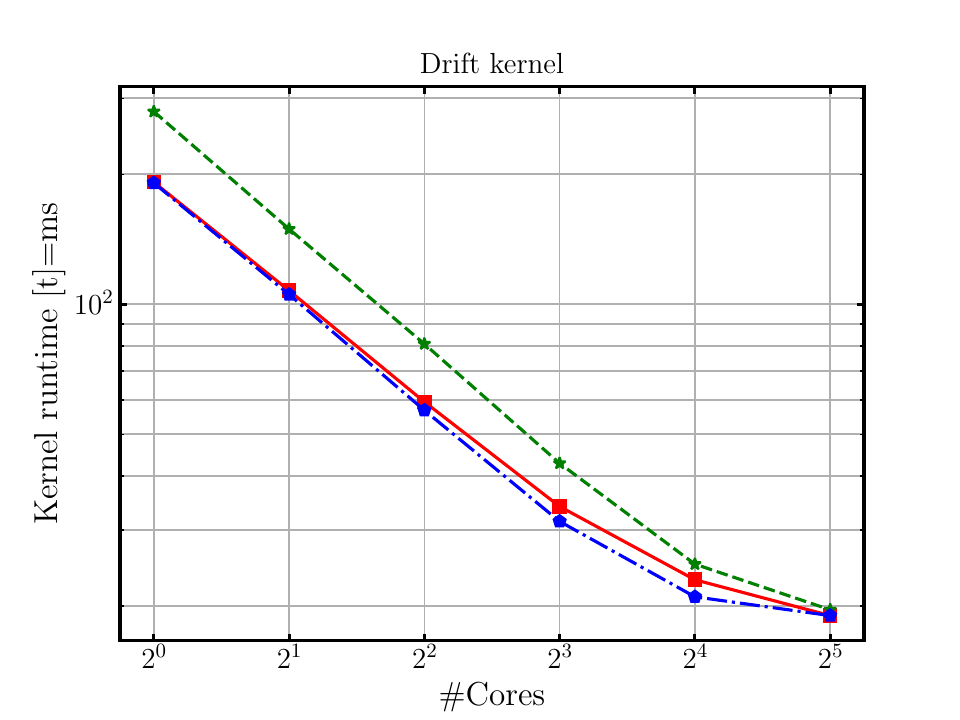}
    \includegraphics[width=0.4\textwidth]{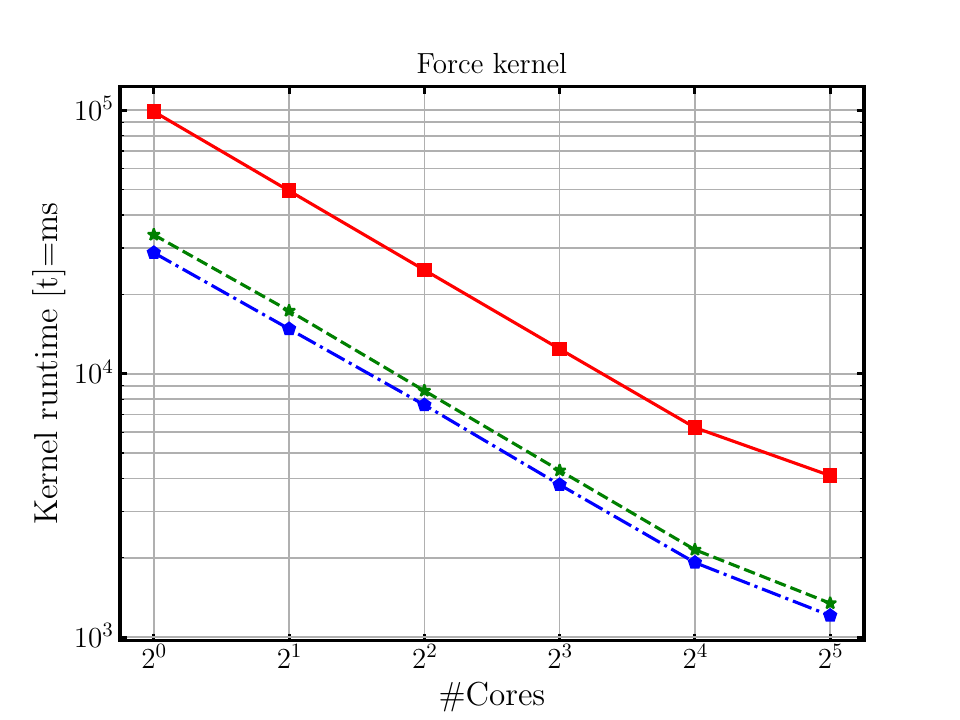}
    \\
    \includegraphics[width=0.4\textwidth]{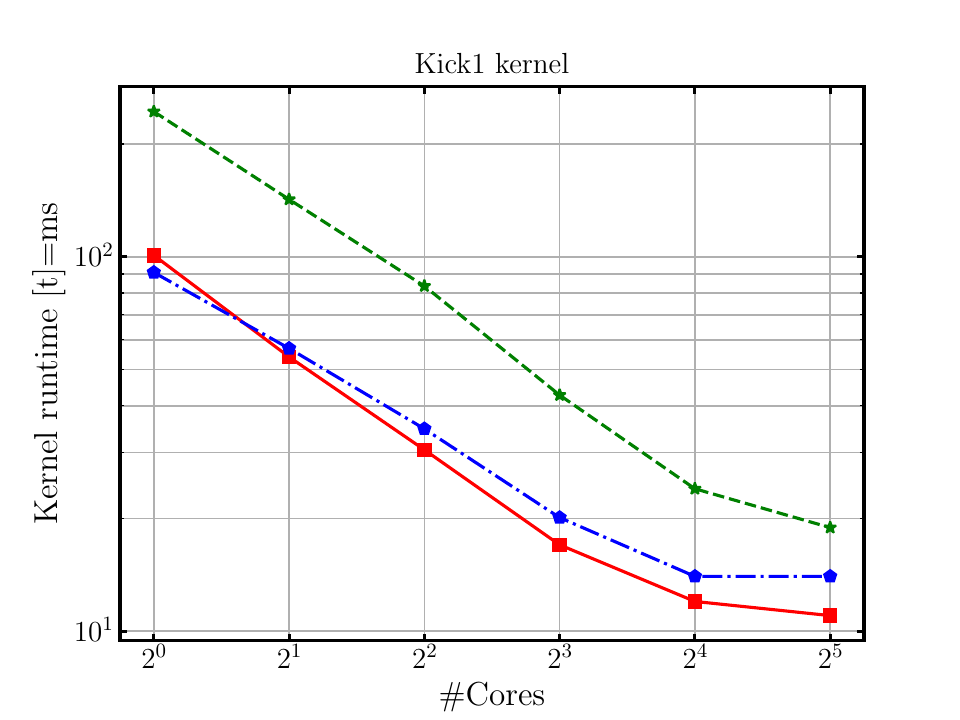}
    \includegraphics[width=0.4\textwidth]{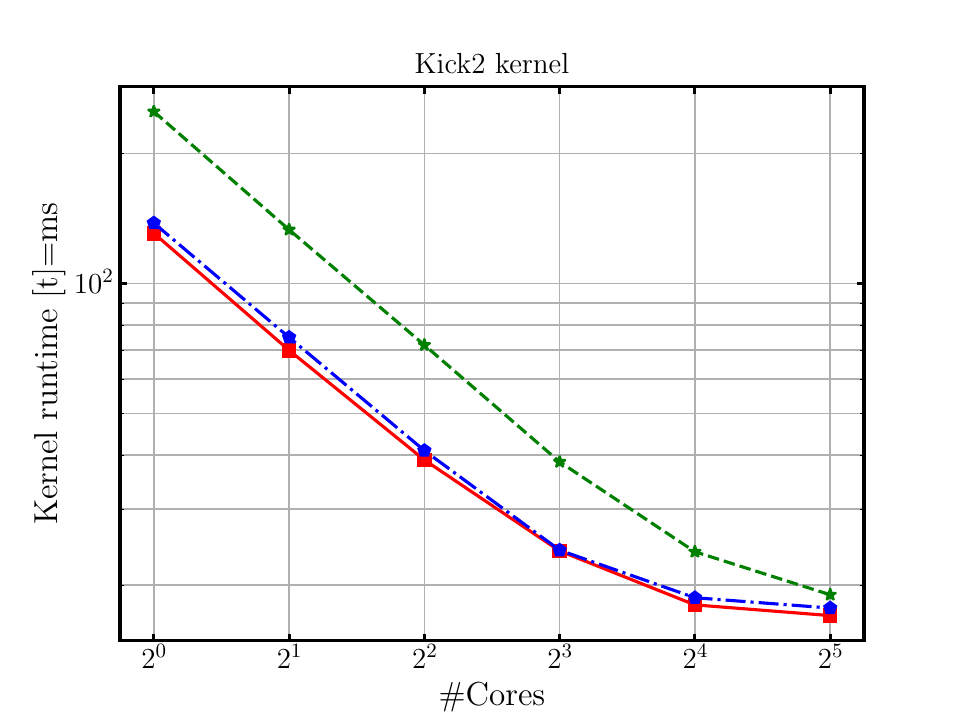}
  \end{center}
    \noindent
    Experiment from Figure~\ref{figure:results:scaling:genoa} for
    $16 \cdot 2048^2$ particles.

\section{Additional throughput data}
\label{appendix:throughput-data}

  \begin{center}
    \includegraphics[width=0.4\textwidth]{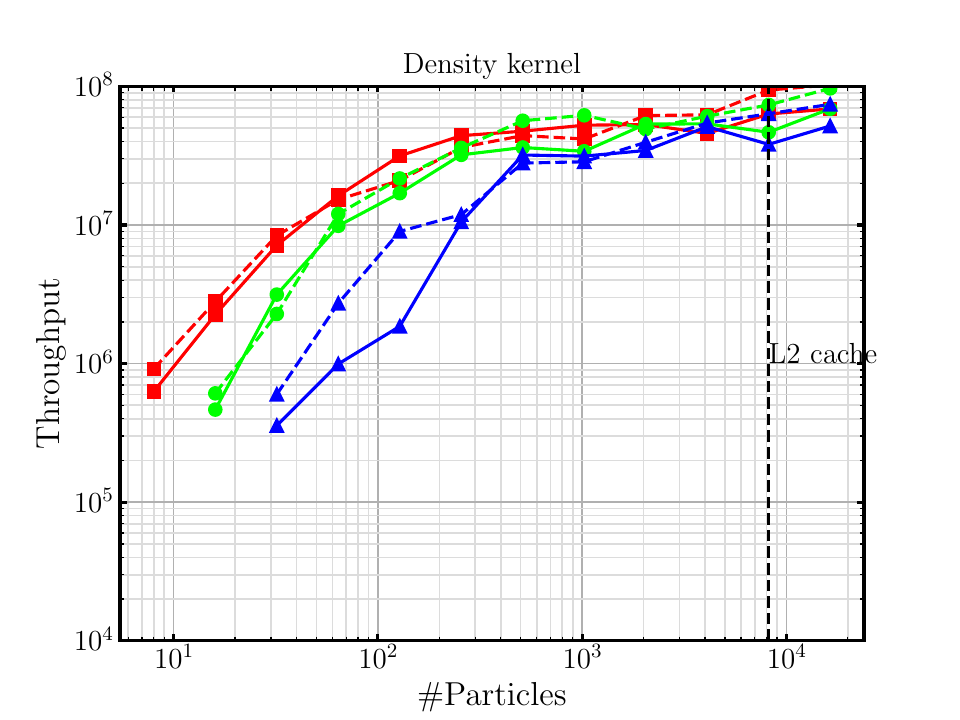}
    \includegraphics[width=0.4\textwidth]{experiments/peano/sapphirerapid/plot-Force.pdf}
  \end{center}
    \noindent
    All Sapphire Rapid data for the algorithmic steps with quadratic complexity.

  \begin{center}
    \includegraphics[width=0.4\textwidth]{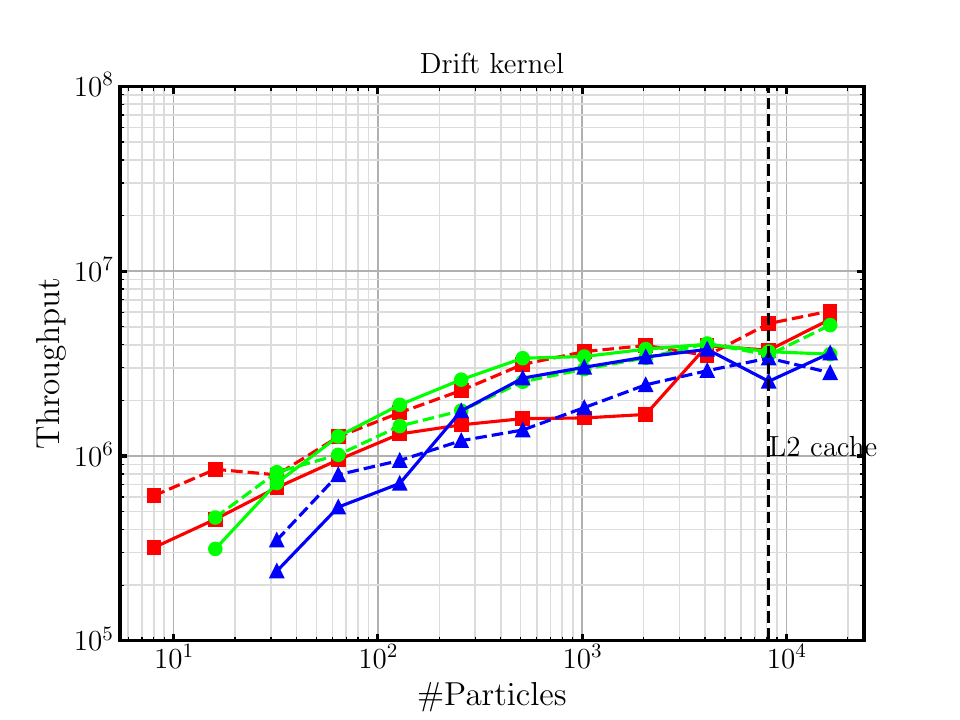}
    \includegraphics[width=0.4\textwidth]{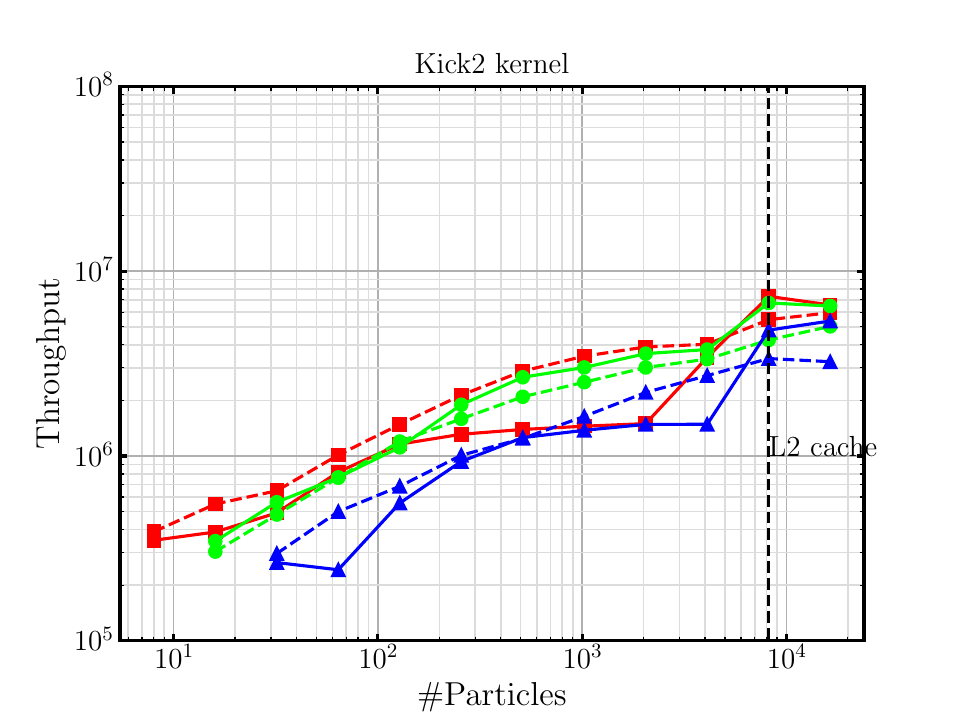}
  \end{center}
    \noindent
    Sapphire Rapid data for the algorithmic steps with linear complexity which
    are not show in main manuscript.

  \begin{center}
    \includegraphics[width=0.4\textwidth]{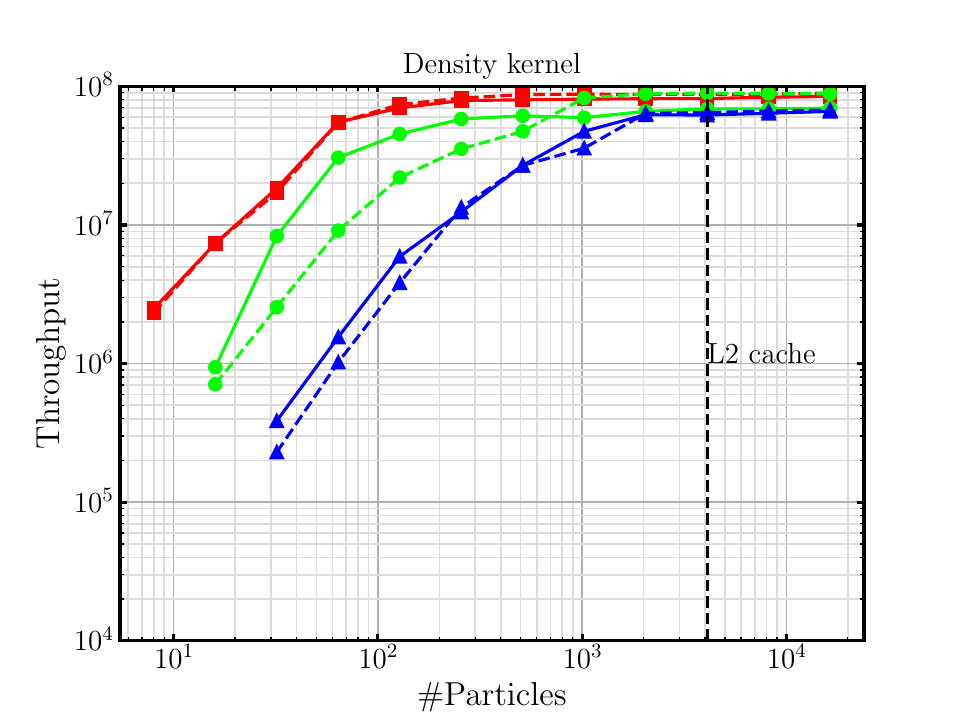}
    \includegraphics[width=0.4\textwidth]{experiments/peano/genoa/plot-Force.pdf}
  \end{center}
    \noindent
    Genoa data for algorithmic steps with quadratic complexity.

  \begin{center}
    \includegraphics[width=0.4\textwidth]{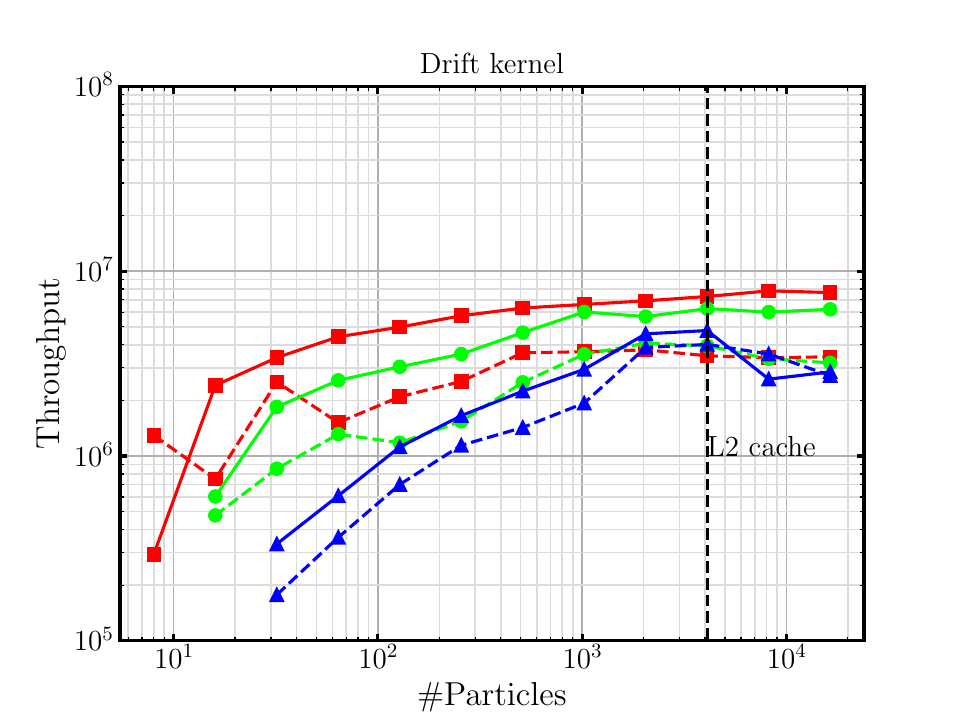}
    \includegraphics[width=0.4\textwidth]{experiments/peano/genoa/plot-Kick1.pdf}
    \includegraphics[width=0.4\textwidth]{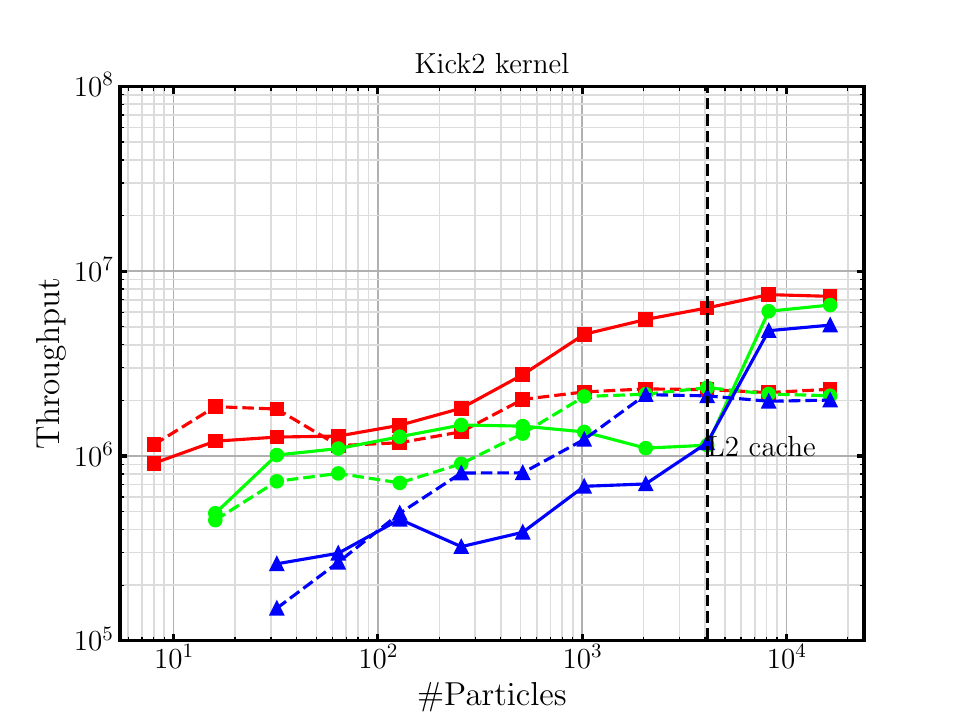}
  \end{center}
    \noindent
    Genoa data for algorithmic steps with linear complexity.

\end{document}